\documentclass{ws-procs975x65}

\begin{document}

\title{Search for $\nu_\mu\rightarrow\nu_\tau$ oscillations in appearance mode in the OPERA experiment}

\author{L. S. Esposito$^*$, on behalf of the OPERA Collaboration}

\address{ETH, Institute for Particle Physics,\\
Zurich, Switzerland\\
$^*$E-mail: luillo@cern.ch}

\begin{abstract}
The OPERA experiment is aiming at the first direct detection of neutrino oscillations in appearance mode through the study of the $\nu_\mu\rightarrow\nu_\tau$ channel. The OPERA detector is placed in the CNGS long baseline $\nu_\mu$ beam 730 km away from the neutrino source. The analysis of a sub-sample of the data taken in the 2008-2009 runs was completed. After a brief description of the beam and the experimental setup, we report on event analysis and on a first candidate event, its background estimation and statistical significance.
\end{abstract}


\bodymatter

\section{Introduction}
\label{intro}
The phenomenon of neutrino oscillation, the transition from a neutrino flavour to another, was anticipated nearly 50 years ago\cite{pontecorvo}. Two types of experimental methods can be used to detect such oscillations: observe the appearance of a neutrino flavour initially absent in the beam or measure the disappearance rate of the initial flavour.
The disappearance of muonic neutrinos have been convincingly observed in different experiments~\cite{SK,MINOS}. In the SNO~\cite{SNO} solar neutrino experiment, the measured rate of neutral-current (NC) interactions was shown to be compatible with the rate expected from total solar neutrino flux, supporting the idea that flavour transitions occur among the three active flavours of the Standard Model. 
However, there is no direct evidence of neutrino oscillation by the appearance method where the new flavour is identified.
The OPERA experiment is aiming at the first direct detection of neutrino oscillations in appearance mode by the identification of the $\tau$ lepton produced in $\nu_\tau$ CC interaction in an almost pure muon neutrino beam by an event-by-event measurement.
In a different approach, a statistical analysis of the atmospheric neutrino 
in Super-Kamiokande shows that data are inconsistent with no $\tau$ appearance hypothesis due to $\nu_\mu\rightarrow\nu_\tau$ oscillations~\cite{SK_tau}. 

\section{Beam}
\label{beam}
OPERA is exposed to the long-baseline CNGS $\nu_\mu$ beam \cite{CNGS}  from CERN in the Gran Sasso Laboratories (LNGS), 730~km from the CERN neutrino source. The beam is optimized for the observation of $\nu_\tau$  CC interactions. The average neutrino energy is $\sim$17 GeV. The $\bar{\nu}_\mu$ contamination is 2.1\% in terms of interactions; the $\nu_e$ and  $\bar{\nu}_e$ contaminations are lower than 1\%, while the number of prompt $\nu_\tau$  is negligible.  
With a total CNGS beam intensity of $22.5 \times 10^{19}$ pot (protons on target), about 24300 neutrino events would be collected. The experiment should observe about 10 $\nu_\tau$~CC events for the present $\Delta m^2_{23}$ allowed region with a background of less than one event.  During the physics runs in 2008, 2009 and 2010 the total achieved intensity was $9.34 \times 10^{19}$~pot.

\section{The detector}
\label{detector}
\begin{figure}[h!]
\begin{center}
  \includegraphics[width=0.7\linewidth]{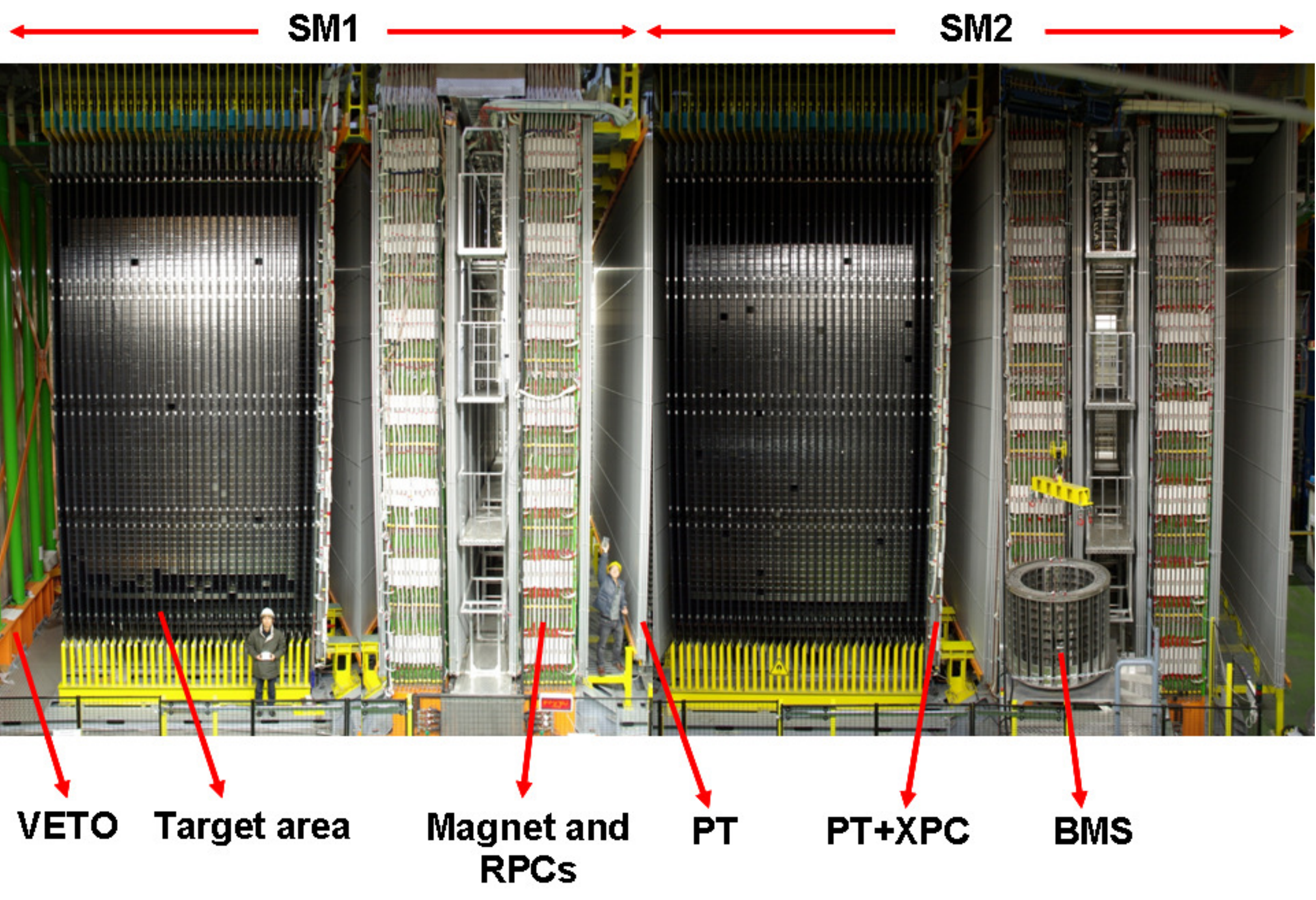}
   \caption{The OPERA detector in the LNGS hall C, at a depth corresponding to 3,100 m water equivalent overburden.
The CNGS neutrino beam comes from left.} \label{fig:det}
\end{center}
\end{figure}

The OPERA detector installed in the underground laboratory of LNGS has three main components (see Fig.~\ref{fig:det}).

\begin{figure}[h!]
\begin{center}
  \includegraphics[width=0.7\linewidth]{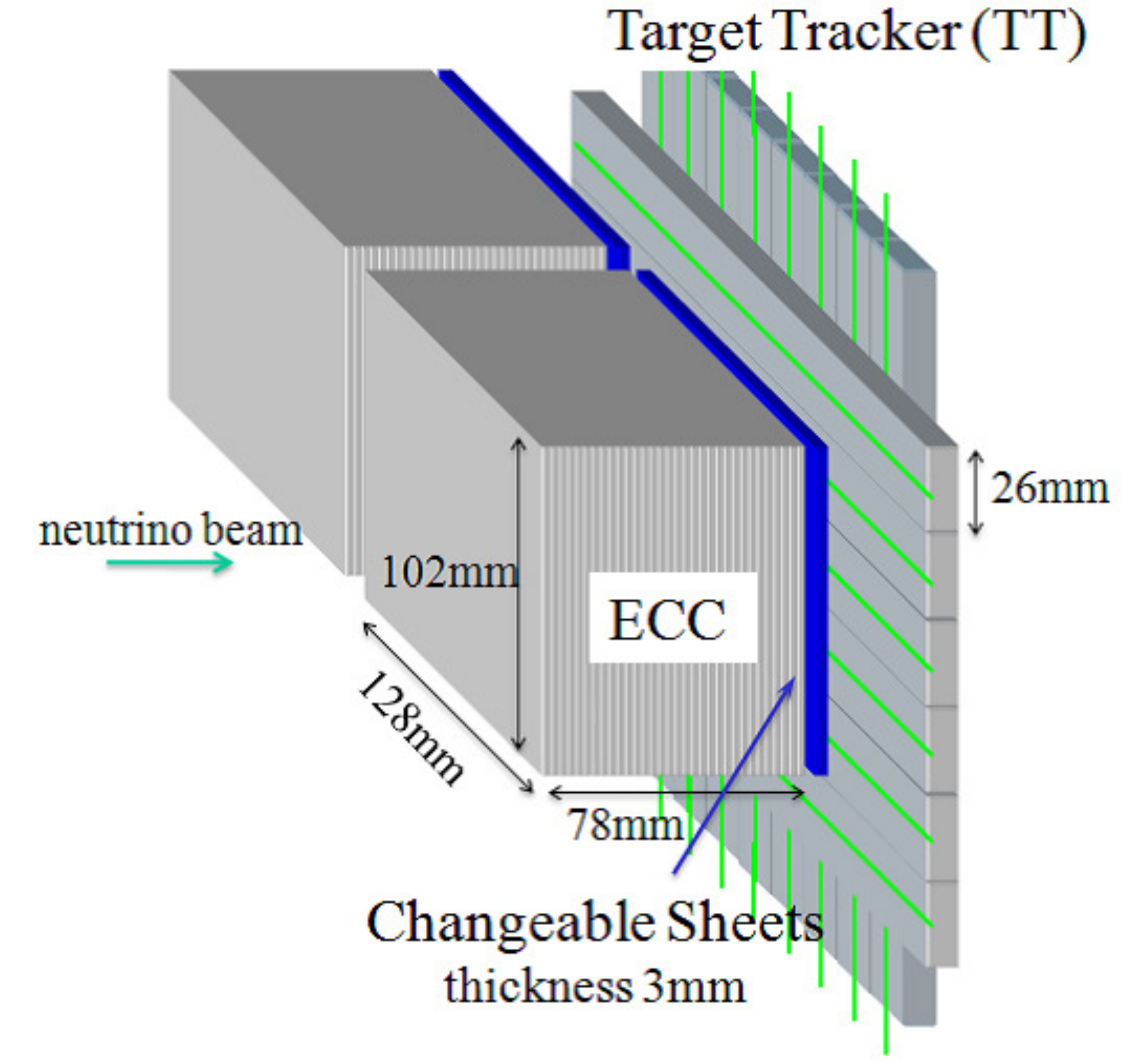}
   \caption{Arrangement of the ECC brick walls and the TT scintillator 
planes.}\label{fig:ecc-tt}
\end{center}
\end{figure}

The first one is an active target, consisting of about 150k Emulsion Cloud Chamber (ECC) modules called ”bricks”. A brick is a sandwich structure of 57 nuclear emulsion films and 56 1 mm thick lead plates. The lead plates serve as neutrino interaction target and the emulsion films as 3-dimensional tracking detector providing track coordinates with a sub-micron accuracy and track angles with a few mrad accuracy. The material of a brick along the beam direction corresponds to about 10 radiation length and 0.33 interaction length. The brick size is 10~cm$\times$12.5~cm$\times$8~cm and it weighs about 8.3~kg. The total active target mass is thus 1.25 ktons. The target is subdivided in two identical units, each consisting of walls of bricks, containing about 2800 units each.
The second main component of the detector is the target tracker (TT) system, a set of 62~scintillator planes in total, interleaved with the brick walls. Each TT plane consists of 256 2.6~cm wide plastic strips. The sandwich structure of the brick walls and the TT planes is illustrated in Fig.~\ref{fig:ecc-tt}.

The third detector component consists in two muon spectrometers following the two target units. Equipped with RPC and high precision drift chambers, they allow determining the muon momentum with better than 20\% accuracy up to 30 GeV/c.

\section{Event analysis}
\label{analysis}
Neutrino events analysis starts with processing of signals from the electronic detectors, reconstructing tracks of particles outgoing from the interaction vertex and measuring muons momentum in the magnetic spectrometers. Tracking information is combined with reconstructed hadronic shower axis and the output of a Neural Network for the selection of the wall where the interaction occurred, providing a list of bricks with the associated probability that the interaction occurred therein. For events with a muon in the final state, a prediction for the slope of the muon and its impact on the brick is provided. For NC events a hadronic shower calculated with TT hits provides general direction to the vertex. This part of the event analysis is called Brick Finding (BF). The brick with the highest probability is extracted from the detector for analysis. No new bricks are inserted in the detector during the experiment, so the total target mass is gradually decreasing with time. Therefore, a high efficiency of the BF is important for minimization of the target mass loss and the reduction of emulsion processing load.
After extraction of the brick predicted in the electronic detectors, its validation is performed by analysis of two interface emulsion films (called Changeable Sheets, CS)~\cite{CS} that are inserted in between each ECC brick and TT scintillator strips. The CS doublet is analysed in the scanning facilities before a brick is disassembled. The information of the CS is then used for a precise prediction of the position of the tracks in the most downstream films of the brick, hence guiding the scan-back vertex finding procedure. If no tracks are found in the CS, the brick is returned back to the detector with another CS doublet attached; 
otherwise,
after a brick has been validated, it is dismantled and its emulsion films are developed and dispatched to the various scanning laboratories, in Europe or in Japan
After a brick has been validated, its emulsion films are developed and dispatched to the various scanning laboratories. All tracks measured in the CS are followed back until they are not found. The stopping point is considered as either a primary or a secondary vertex, then the vertex is confirmed by scanning a volume with 1~cm$^2$ in 15 films, 5 upstream and 10 downstream of the stopping point. A further analysis called decay search procedure is applied to located vertices to detect possible decay or interaction topologies. When secondary vertices are found in the event, a kinematical analysis is performed using track angles and momenta.
Momenta of charged particles can be measured in ECC using the angular deviations of tracks by Multiple Coulomb Scattering (MCS) in lead~\cite{MCS}. This method gives a momentum resolution better than 22\% for particles with momenta lower than 6 GeV/c, passing through a brick. For higher momentum, the position deviations are used for the measurement. The resolution is better than 33\% for particles with momenta lower than 12 GeV/c, passing through a brick. Momenta of muons reaching the spectrometer are measured with a resolution of 20\% up to 30 GeV/c.

The analysis of a sub-sample of 1088 events of the neutrino data taken in the 2008-2009 runs was completed, corresponding to $1.89 \times 10^{19}$ pot.
Charmed particles have similar lifetimes and decay topologies if charged, so the detection of charm decays is used to check the $\tau$ detection efficiency. In the sample of $\nu_\mu$~CC interactions, 20 charm decays have been observed that survived all the cuts, in agreement with expectations from a MC study, $16.0\pm 2.9$. Out of them 3 have a 1-prong topology where $0.8\pm0.2$ are expected. The background for the total charm sample is about 2 events.
Several $\nu_e$-induced events have also been observed.

\section{The first $\nu_\tau$ candidate event}
\label{tau}

\begin{figure}[h!]
\begin{center}
  \includegraphics[width=0.445\linewidth]{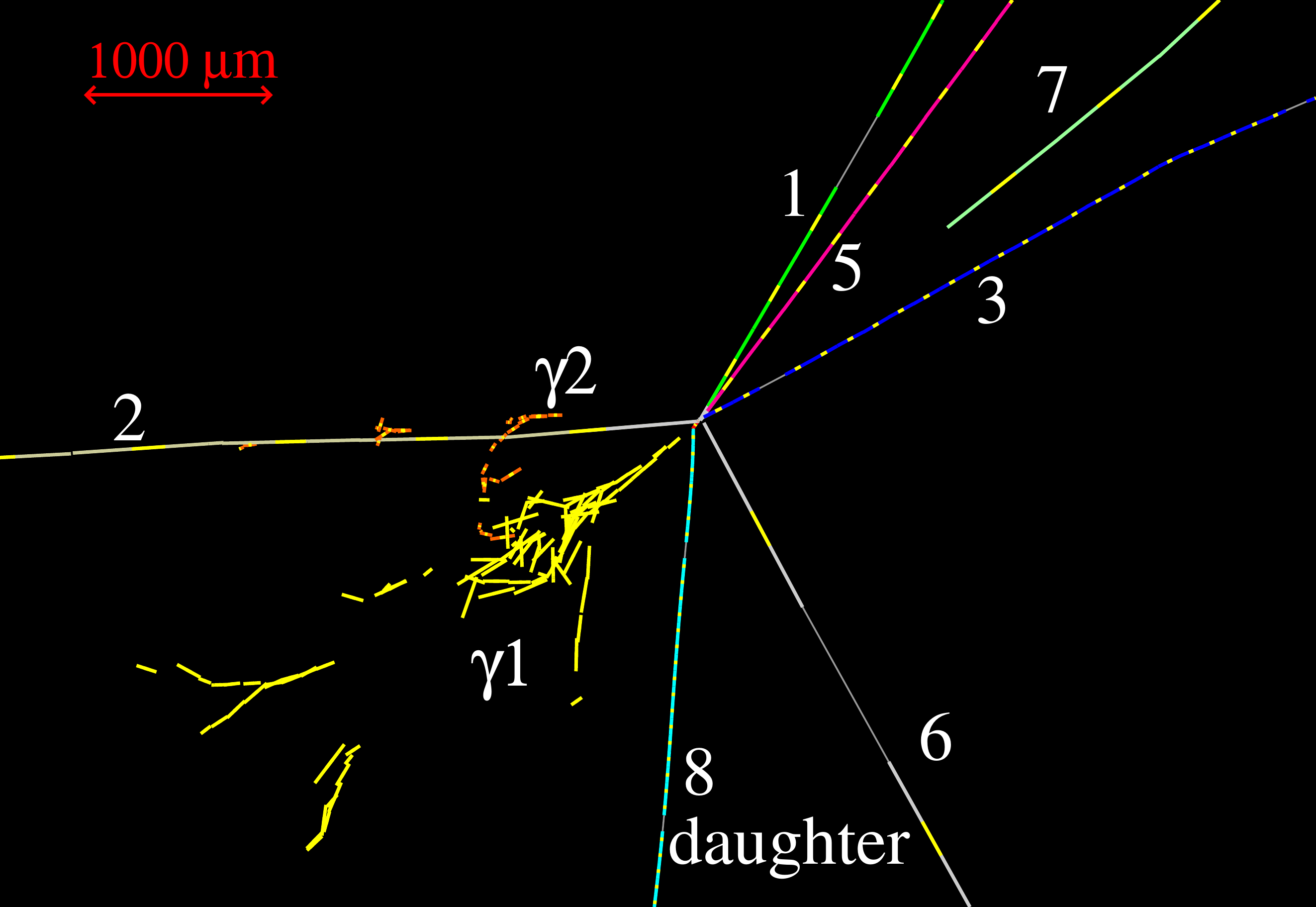}
  \includegraphics[width=0.445\linewidth]{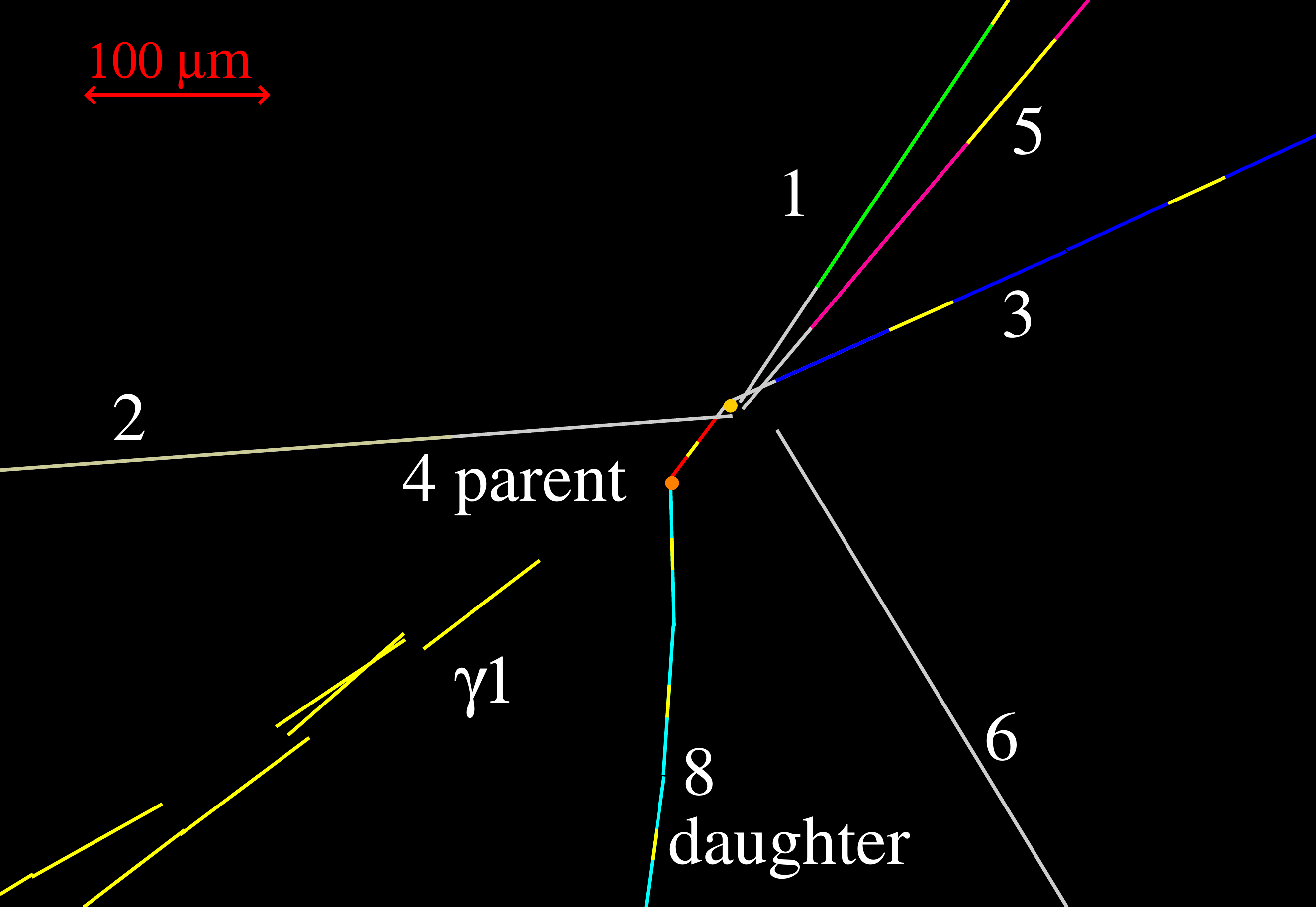}\\
  \includegraphics[width=0.9\linewidth]{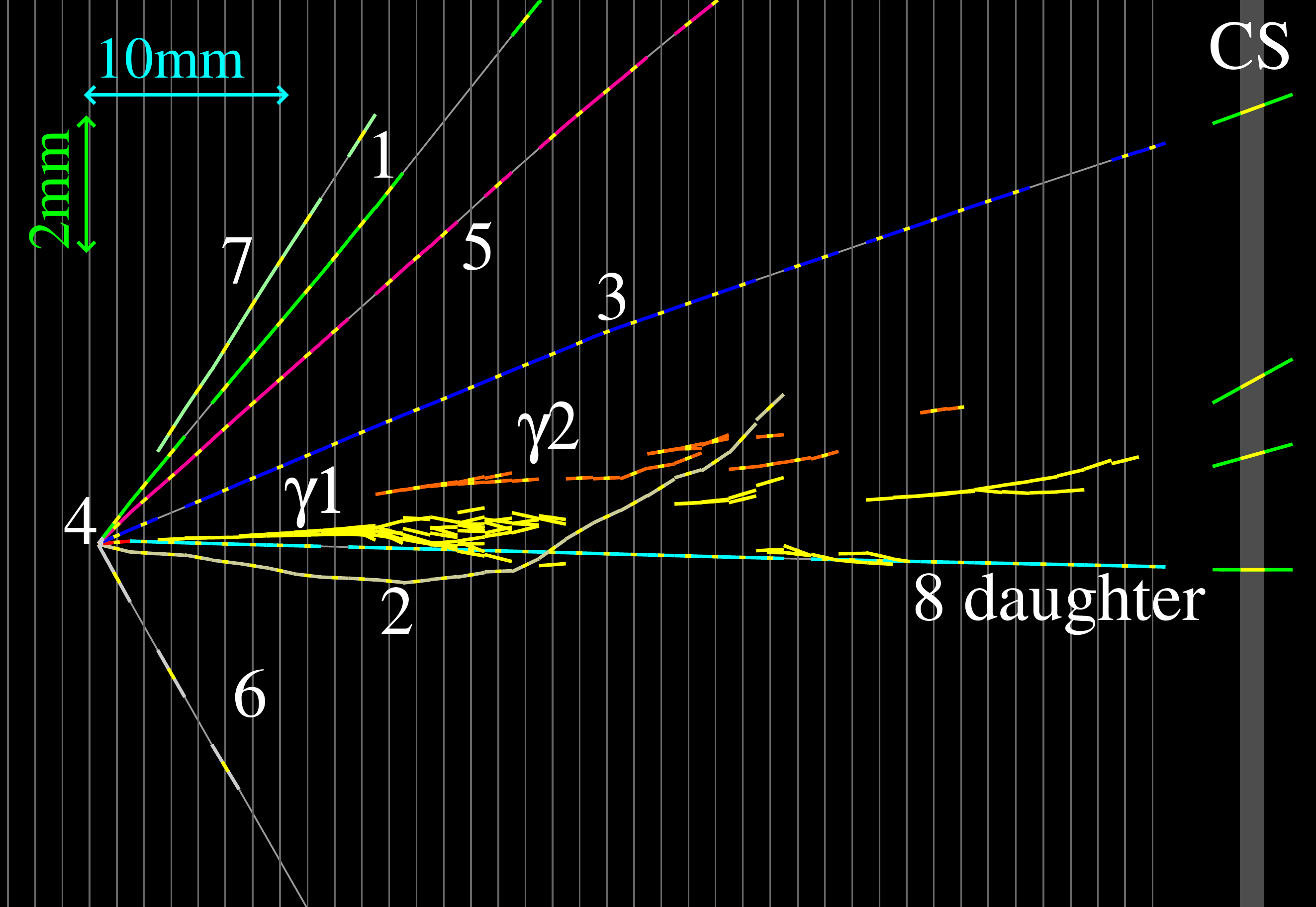}
   \caption{Display of the $\tau^-$ candidate event. {\it Top left:} view transverse to the neutrino direction. {\it Top right:} same view zoomed on the vertices. 
{\it Bottom:} longitudinal view.} \label{event}
\end{center}
\end{figure}

The decay search procedure yielded one candidate event satisfying the selection criteria for the $\nu_\tau$ interaction search~\cite{OPERA_tau}. The cuts are the same as those defined in the experimental proposal~\cite{OPERA}. The event is displayed in Fig.~\ref{event}.

The selection criteria are that there are no primary tracks compatible with a muon or an electron, and the secondary vertex and the primary vertex survive the cuts shown in Table~\ref{tab:var}.

\begin{table}
\tbl{Selection criteria values for the $\nu_\tau$ candidate event.}
{\begin{tabular}{@{}ccc@{}}
\toprule
Variable         & Measured & Selection criterion \\
\colrule
Kink angle     & $42\pm2$ & $>20$ \\
(mrad) & & \\
Decay length & $1335\pm35$ & within 2 plates \\
($\mu$m) & & \\
$P$ daughter     & $12^{+6}_{-3}$ & $>2$\\
(GeV/c) & & \\
$P_{t}$ daughter& $470^{+230}_{-120}$ & $>300$ ($\gamma$ attached)\\
(MeV/c) & & \\
Missing $P_{t}$  & $570^{+320}_{-170}$ & $<1000$\\
(MeV/c) & & \\
Angle $\Phi$ & $173 \pm 2$ & $>90$\\
(deg) & & \\
\botrule
\end{tabular}
}
\label{tab:var}
\end{table}

The primary vertex consists of seven tracks of which one track has a kink. Two electromagnetic showers caused by $\gamma$~rays have been detected and they are compatible with pointing to the secondary vertex.
The invariant mass of the two detected $\gamma$~rays yields a mass consistent with	the $\pi^{0}$~mass, $(120\pm20~(stat) \pm 35~(syst))$~MeV/c$^{2}$. Then, the invariant mass of the $\pi^{-}\gamma\gamma$ system has a value compatible with that of the 
$\rho (770)$, $(640 \pm 125 ~(stat) \pm 100~(syst)$~MeV/c$^{2}$.
The $\rho$ appears in about 25\% of the $\tau$ decays.

\section{Background estimation and statistical significance}
\label{background}

The secondary vertex is compatible with the decay of $\tau \rightarrow h^{-} (n \pi^{0}) \nu_\tau$. The main background sources to this channel are
\begin{itemize}
\item the decays of charmed particles produced in $\nu_\mu$~CC interactions where the primary muon is not identified.
\item the 1-prong interactions of primary hadrons produced in $\nu_\mu$~CC interactions where the primary muon is not identified or in $\nu_\mu$~NC interactions.
\end{itemize}

The charm background produced in the $\nu_\mu$ interactions in the analyzed sample is $0.007\pm0.004$~events, that produced in the $\nu_e$ interactions is less than $10^{-3}$ events.

The background from hadron interactions has been evaluated with a FLUKA based MC code, updated with respect to the proposal simulations. The kink probability to occur in 2~mm lead integrated over the $\nu_\mu$~NC hadronic spectrum yield a background probability of
$(1.9 \pm 0.1) \times 10^{-4} $/NC. This probability decreases to 
$(3.8 \pm 0.2) \times 10^{-5} $/NC taking into account the cuts on the event global kinematics. This leads to a total of $0.011\pm0.006$~events when misclassified CC events are included.
Cross-checks of the 1-prong hadron background estimation have been performed. The tracks of hadrons from a sub-sample of neutrino interactions have been followed far from the primary vertex to search decay-like interactions. A total length of 9 m has been measured and no event has been found in the signal region. This corresponds to a probability over 2 mm lead smaller than $1.54 \times 10^{-3}$ at 90\% C.L. 
Within the low statistics, the $P_t$ distribution agrees with the simulation.

In summary  we observed 1 event in the 1-prong hadron $\tau$ decay channel, with a background expectation $0.018\pm0.007~(syst)$. The probability to observe 1 event due to a background fluctuation is 1.8\% (2.36$\sigma$). As all $\tau$ decay modes were included in the search (1-prong $\mu$, 1-prong e, 1-prong hadron, 3-prong hadron), the total background becomes 
$0.045 \pm 0.023 ~(syst)$, the probability to observe 1 event due to a background fluctuation becomes 4.5\% ($2.01 \sigma$).

\section{Conclusion}
\label{conclusion}

Data taking in the CNGS beam is going smoothly. The analysis of a sub-sample of the neutrino data taken in the 2008-2009 runs was completed, corresponding to \mbox{$1.89 \times 10^{19}$~pot} out of $22.5 \times 10^{19}$ proposed pot. Decay topologies due to charmed particles have been observed in good agreement with expectations, as well as several events induced by $\nu_e$ present as a contamination in the $\nu_\mu$ beam. 
One muon-less event showing a $\tau$ to 1-prong hadronic decay topology has been detected. It is the first
$\nu_\tau$ candidate event in OPERA, with a statistical significance of $2.36 \sigma$ 
(1-prong hadronic decay mode) and $2.01 \sigma$ (all decay modes). Analysis of the 2008+2009 full sample will be completed in two months.
Analysis of 2010 events is being performed in parallel.
Next CNGS runs will be in 2011 and 2012 (CERN stop in 2013).

\bibliographystyle{ws-procs975x65}

\end{document}